\documentclass[notitlepage,reprint,superscriptaddress,longbibliography,nofootinbib]{revtex4-1}

\usepackage{graphics}
\usepackage{dcolumn}
\usepackage{bm}
\usepackage[caption=false]{subfig}
\usepackage{xfrac}
\usepackage{mathrsfs}
\usepackage{float}
\usepackage[normalem]{ulem}
\usepackage[bbgreekl]{mathbbol}
\usepackage[utf8]{inputenc}
\usepackage{amsmath} 
\usepackage{multirow}
\usepackage{adjustbox}
\usepackage{etoolbox}
\usepackage{comment}
\usepackage{amsmath}
\usepackage{amssymb}
\usepackage{blindtext}
\usepackage[colorlinks]{hyperref}

\usepackage{graphicx}
\usepackage{capt-of}

\newcommand{\ket}[1]{|{#1}\rangle}
\newcommand{\bra}[1]{\langle{#1}|}
\newcommand{\bracket}[2]{\langle{#1}|{#2}\rangle}

\newcommand{\figref}[1]{Fig.~\ref{fig:#1}}
\newcommand{\figrefbegin}[1]{Figure~\ref{fig:#1}}

\renewcommand{\eqref}[1]{Eq.~(\ref{eq:#1})}

\newcommand{\citeasnoun}[1]{Ref.~\citenum{#1}}

\newcommand{\citeasnounss}[1]{Refs.~\citenum{#1}}

\newcommand{\Scale}[2][4]{\scalebox{#1}{$#2$}}%

\begin{document}
\author{A. Pick}
\email{pick.adi@gmail.com}
\affiliation{Faculty of Chemistry, Technion-Israel Institute of Technology, Haifa, Israel.}
\affiliation{Faculty of Electrical Engineering, Technion-Israel Institute of Technology, Haifa, Israel.}
\author{S. Silberstein}
\affiliation{Department of Physics,  Hebrew University, Jerusalem 9190401, Israel}
\author{N. Moiseyev}
\affiliation{Faculty of Chemistry, Technion-Israel Institute of Technology, Haifa, Israel.}
\affiliation{Faculty of Physics, Technion-Israel Institute of Technology, Haifa, Israel.}
\author{N. Bar--Gill}
\affiliation{Department of Physics,  Hebrew University, Jerusalem 9190401, Israel}
\affiliation{Department of Applied Physics,  Hebrew University, Jerusalem 9190401, Israel}

\title{Robust  mode conversion  in  NV centers using exceptional points}

\begin{abstract}
We show that microwave-driven NV centers  can function as robust  mode switches by utilizing a special degeneracy called an exceptional point (EP).   While previous theoretical and experimental work on EP-based mode switches applies only to pure states, we develop here a general theory for switching between mixed states---statistical ensembles of different pure states, resulting from the interaction with the environment. Our  theory is general and applicable to  all  leading platforms for quantum information processing and quantum technologies. However, our numerical simulations use empirical parameters of  NV centers. We provide guidelines for coping with the main challenges for experimental realization of this protocol: decoherence and mixed-state preparation. \end{abstract}

\maketitle

A new class of adiabatic protocols  enables robust mode conversion in open systems that possess a special  degeneracy called  an exceptional point (EP)---where multiple modes of the system coalesce~\cite{kato2013perturbation,strang2006linear,trefethen2005spectra}.  EP-based mode  switches  have   intriguing physical properties, such as topological protection and non-reciprocity~\cite{uzdin2011observability,berry2011slow,milburn2015general,hassan2017chiral,hassan2017dynamically}, which were    demonstrated experimentally in optical waveguides and optomechanics~\cite{doppler2016dynamically,xu2016topological,yoon2018time,zhang2018dynamically,zhang2018dynamicallyB,zhang2018hybrid} and theoretically  proposed for  several  additional systems~\cite{chatzidimitriou2018optical,kapralova2014helium,gilary2013time,ke2016exceptional}.  
Realizing robust non-reciprocal mode switching in quantum systems has far reaching consequences in quantum information processing and quantum control, as well as in quantum technology. Here we show how to realize EP-based mode  switches in atomic and atom-like systems.  While previous work on EP-based mode switches applies only to pure states,  the theoretical description of  atom-like  systems  typically requires  mixed states---statistical ensembles of different pure states, which arise due to interactions with the environment. To bridge this gap, we develop  a theory of mode switching between mixed states. Our protocol applies, most generally,    to  three-level systems in the V-configuration, and we perform numerical simulations using empirical parameters of  nitrogen-vacancy (NV) centers---defects in  diamond   with exceedingly long coherence lifetimes  and established  optical and microwave mechanisms for initialization, manipulation and readout of their spin state~\cite{taylor2008high,maze2008nanoscale,balasubramanian2008nanoscale,jelezko2006single}.   
  Our theory enables exploring new phenomena (e.g., high-order EPs in low-dimensional systems) and presents a crucial step towards incorporating   EP-based mode switches  in quantum-information applications.

Robust   mode switches are    based on the adiabatic theorem, which describes the evolution of slowly varying   \emph{closed systems}. The theorem states that when preparing a  system  in a  particular   eigenmode, it remains in that mode during  the evolution (given the conditions specified in~\citeasnounss{born1928beweis,kato1950adiabatic}). Dynamic closed systems are  described by Hermitian Hamiltonians that  depend on  a set of  ``control parameters.''  When changing the  parameters slowly along  closed loops in parameter space,   the  theorem implies that   the  initial and final  states are  the same  (up to a     phase~\cite{pancharatnam1956generalized,longuet1958studies,berry1984quantal}). The  understanding  that topological operations (i.e., executing closed control paths)  may have  outcomes that  are  robust against noise has  lead to important discoveries in  multiple areas of physics~\cite{xiao2010berry,nayak2008non,lu2014topological,mead1992geometric}.

  \begin{figure*}[t]
  \begin{center}
    \includegraphics[scale=0.5]{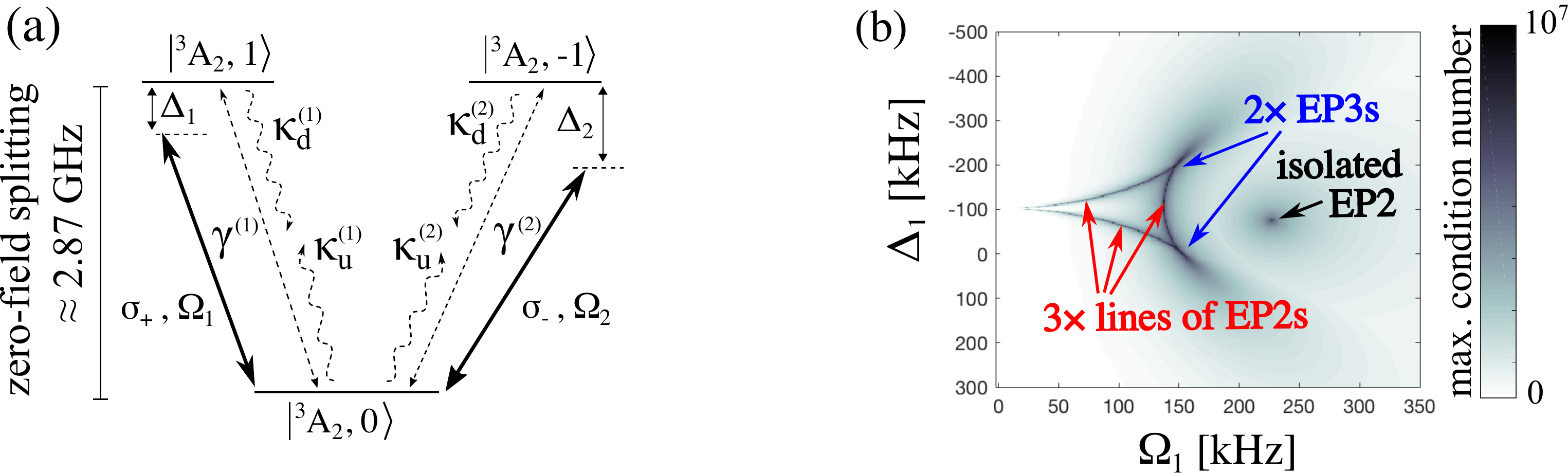}
  \end{center}
  \caption{\textbf{(a) Ground   states of the  NV center.} 
  Two    microwave fields  drive transitions  between the $\ket{^3\!A_2, 0}$ and $\ket{^3\!A_2,\pm1}$  states, with 
Rabi frequencies $\Omega_{1,2}$,  detunings   $\Delta_{1,2}$,  and  left-/right-circular  polarizations $\sigma_\pm$. Interaction with the  environment causes  pure dephasing at a rate  $\gamma^{(1,2)}$ (straight dashed arrows) and 
 downwards or upwards  transitions at  rates $\kappa_u^{(1,2)}$ or $\kappa_d^{(1,2)}$  respectively (wiggly dashed arrows).
  The  evolution of the system is governed by the  Lindbladian operator $\hat{\mathcal{L}}$ [\eqref{Lindblad}].
 \textbf{(b) Exceptional points in the Lindbladian.}
A  filled contour plot of  the maximal condition number of the eigenvalues of  $\hat{\mathcal{L}}$ [\eqref{condition-num}]. We scan $\Delta_1$ and $\Omega_1$ while fixing all other parameters: $\Omega_2 = 400, \Delta_2 = 1400,\kappa_u^{(1,2)} = \kappa_d^{(1,2)}= 1, \gamma^{(1)} = 900$, and $\gamma^{(2)} = 1500$ (in  kHz). The condition number diverges when modes coalesce at exceptional points (EPs) (scale  on the right). 
  }
  \label{fig:level-structure}
\end{figure*}

However,   most physical  systems exchange energy or particles with their environment. Open systems can be   described  by effective non-Hermitian operators, and their adiabatic transport properties are drastically  different from closed  ones. 
First, unlike Hermitian operators,   non-Hermitian operators may have EPs,  where multiple  eigenmodes have the same eigenvalue and eigenvector~\cite{kato2013perturbation,strang2006linear,trefethen2005spectra}. 
Near EPs, the eigenvalues are  a multivalued functions of the system's parameters and, consequently,  when changing the  control parameters along  \emph{any closed loop  that encircles an EP}, the  eigenmodes at the initial and final points may differ~\cite{dembowski2001experimental}. For example, when the EP is formed by the coalescence of two modes   (hereafter called an EP2),  the     eigenmodes  swap; i.e.,  $\ket{1}\!\!\rightarrow\!\!\ket{2}$ and $\ket{2}\!\!\rightarrow\!\!\ket{1}$. Secondly,  in  non-Hermitian systems, the adiabatic theorem   holds only  for 
``least decaying states''~\cite{nenciu1992adiabatic,sarandy2005adiabatic,miniatura1990geometrical}, which are eigenmodes $\ell$  with eigenvalues $\lambda_\ell$ whose accumulated decay rate  is positive:
\begin{equation}
\Gamma_\ell(t) \equiv\int_0^t \mathrm{Re}\left[\lambda_\ell(t')-\lambda_j(t')\right]dt'>0, \quad\forall\, j\neq\ell,
\label{eq:gamma-ell}
\end{equation}
for all $t$ during the evolution (see  appendix). 
When only two modes are involved, the accumulated decay rates, $\Gamma_{1,2}(t)$, have opposite signs, which implies that while one  of the states  can evolve adiabatically, the other one  cannot. Therefore,     when  encircling an EP2 along a certain path, either      $\ket{1}\!\!\rightarrow\!\!\ket{2}$   or $\ket{2}\!\!\rightarrow\!\!\ket{1}$~\cite{uzdin2011observability,berry2011slow}.
For any given loop, when reversing  the direction of the path, the sign of the accumulated decay rate is reversed [since  the order of  integration limits in~\eqref{gamma-ell} is exchanged], 
and this is the  source for ``non-reciprocity'' of EP-based switches~\cite{doppler2016dynamically,xu2016topological}.

\begin{figure*}[t]
  \begin{center}
    \includegraphics[scale=0.7]{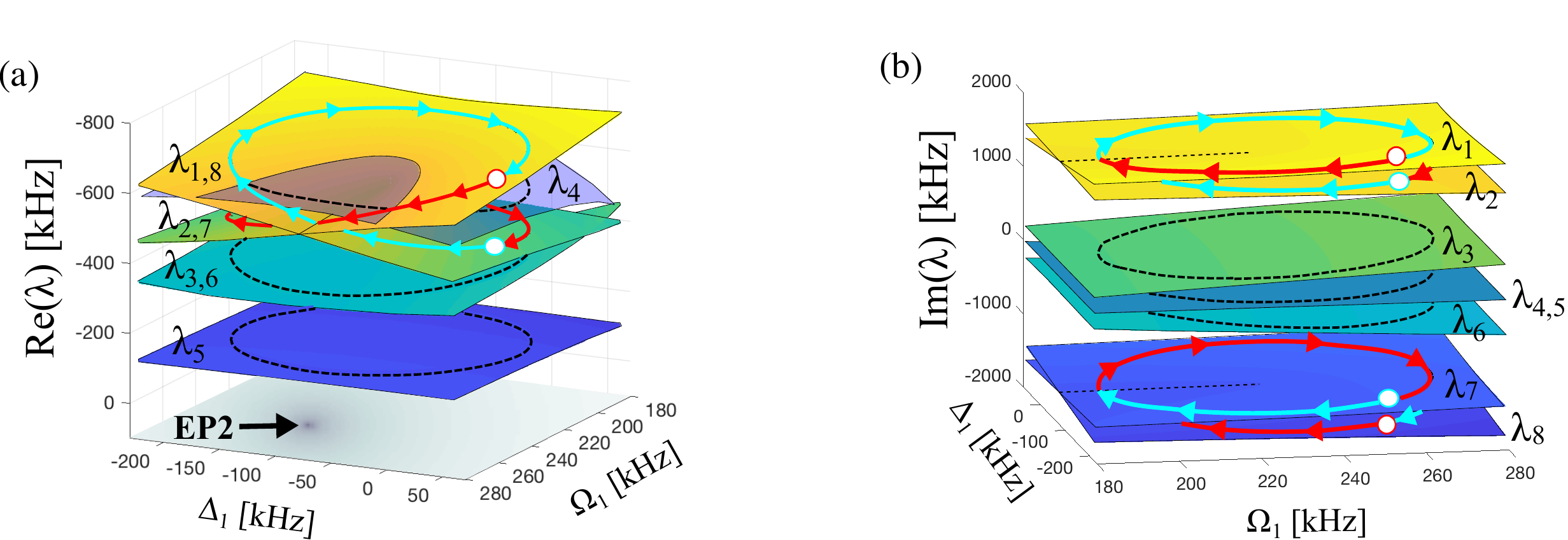}
  \end{center}
  \caption{\textbf{Real and imaginary parts of the eigenvalues of the Lindbladian operator near the isolated EP.}
  We scan   $\Delta_1$ and $\Omega_1$   
   while holding all other parameters  fixed (given  in~\figref{level-structure}).
  In this  regime, the Lindbladian has four simple eigenvalues ($\lambda_{3,4,5,6}$) and two complex-conjugate  pairs    ($\lambda_{1,2}$ and $\lambda_{7,8}$) that coalesce at  the   EP2. 
When changing the  parameters in  a  loop around the EP2 (details in appendix), the simple  eigenvalues  return to the starting point (dashed lines) while the remaining eigenstates swap (arrowed lines); i.e., schematically,  $\ket{1}\leftrightarrow\ket{2}$ and $\ket{7}\leftrightarrow\ket{8}$. 
The red-rimmed points  (associated with $\lambda_{1,8}$) are transferred  into  cyan-rimmed points ($\lambda_{2,7}$) and vice versa.  
}
  \label{fig:riemann-sheets}
\end{figure*}

\begin{figure}[t]
  \begin{center}
    \includegraphics[scale=0.5]{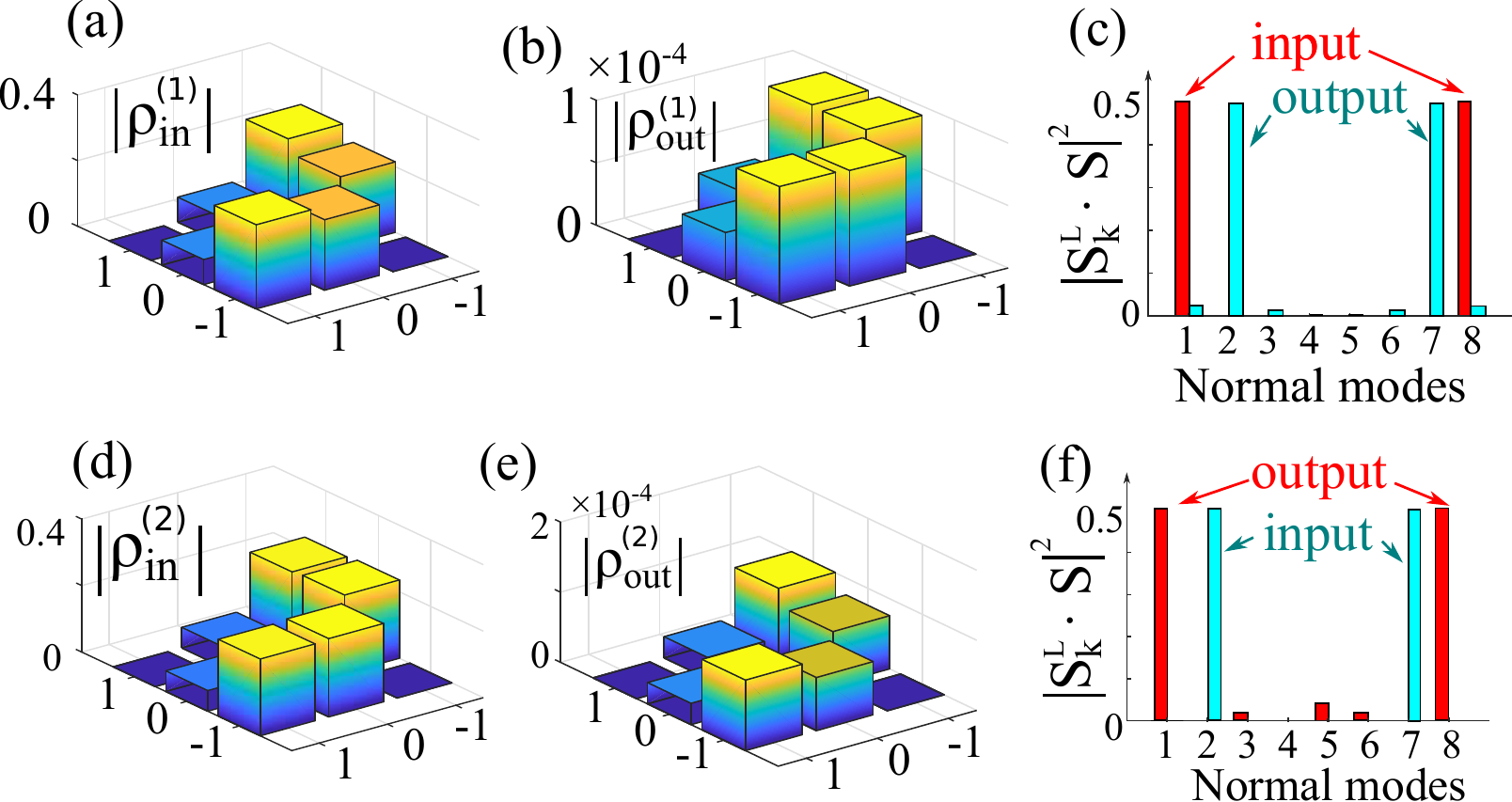}
  \end{center}
  \caption{\textbf{Non-reciprocal  mode switches in NV centers.}
  We initialize the system in either $\hat{\rho}_\mathrm{in}^{(1)}$ or  $\hat{\rho}_\mathrm{in}^{(2)}$ and let it evolve as the parameters are varied along the path from \figref{riemann-sheets}.
Panels (a,b,d,e) show  the  moduli  of  the   off-diagonal density-matrix elements at the input and output. (Diagonal entries  are set to zero for  clarity.)
(a--b) The initial state $\hat{\rho}_\mathrm{in}^{(1)}$  is adiabatically transported into  $\hat{\rho}_\mathrm{out}^{(1)}\propto \hat{\rho}_\mathrm{in}^{(2)}$ after
a clockwise loop around the  EP.
 (d--e) The initial state $\hat{\rho}_\mathrm{in}^{(2)}$ is adiabatically transported into $\hat{\rho}_\mathrm{out}^{(2)}\propto \hat{\rho}_\mathrm{in}^{(1)}$  after a counterclockwise loop. 
(c,f) Normalized projections of the initial and final states on the  basis states  at $t=0$.}
  \label{fig:topo-switch}
\end{figure}

The  performance of the switch  is robust since  it is related to a topological property of the system's energy surfaces---the existence of an EP, which is a  branch point in the system's energy surfaces~\cite{doppler2016dynamically,xu2016topological}. In the adiabatic limit, the output of the switch depends only  on whether or not the loop  encircles an  EP.   Typically, small    perturbations   in the operational details    only slightly distort the loop or move the location of the EP, but  do  not lift the degeneracy. 
 The possibility of creating a  non-reciprocal switches  between pure states has attracted considerable attention~\cite{uzdin2011observability,berry2011slow,milburn2015general,hassan2017chiral,hassan2017dynamically,doppler2016dynamically,xu2016topological,yoon2018time,zhang2018dynamically,zhang2018dynamicallyB,zhang2018hybrid,chatzidimitriou2018optical,kapralova2014helium,gilary2013time,ke2016exceptional} and, here,  we generalize this concept for  mixed states.  
Our protocol includes  
($i$)  finding    an  isolated EP   in the eigenvalue spectrum of the NV center   (\figref{level-structure}),
 ($ii$)  initializing the system in special superposition   states  (\figref{riemann-sheets}), and 
 ($iii$)  changing the  control  parameters in a  loop around the EP   (\figref{topo-switch}).

\figrefbegin{level-structure}(a) shows the electronic ground-state manifold of the negatively charged NV center (also called NV$^-$).  It consists of three spin-triplet   states,    $\ket{^3\!A_2,m_s}$, where $m_s = 0, \pm1$ denotes  the spin projection along the NV axis  and   $A_2$ marks  the orbital symmetry~\cite{doherty2013nitrogen,chu2015quantum}.  In the absence of external magnetic fields,  the energy levels of $\ket{^3\!A_2, 0}$ and $\ket{^3\!A_2,\pm1}$ are split by $2.87$ GHz.
We introduce  right- and left-circularly polarized transverse microwave fields to selectively drive the  $\ket{^3\!A_2, 0}\leftrightarrow\ket{^3\!A_2,\pm1}$  transitions\footnote{Alternatively,
by applying a longitudinal magnetic field to  lift the degeneracy of the  $\ket{^3\!A_2,\pm1}$ states~\cite{togan2011laser}, 
one can   excite these states selectively with a linearly polarized microwave field.}~\cite{alegre2007polarization,london2014strong,mrozek2015circularly} (solid  arrows).  
Using  a semiclassical description~\cite{haken1985laser} (where the electrons are treated quantum mechanically and the   fields are treated classically) and employing the rotating-wave approximation~\cite{scully1999quantum} (which is valid when the driving fields are nearly resonant and  relatively weak), the Hamiltonian of the driven  NV center  in the ground-state manifold   is~\cite{scully1999quantum} 
\begin{align}
&\hat{H} =\Delta_1\ket{1}\bra{1}+\Delta_2\ket{-1}\bra{-1}+\nonumber\\
&-\hbar\Omega_1\left(\ket{1}\bra{0}+\ket{0}\bra{1}\right)
-\hbar\Omega_2\left(\ket{-1}\bra{0}+\ket{0}\bra{-1}\right),
\label{eq:Hamiltonian}
\end{align}
where $\Omega_i\equiv \frac{\mathcal{E}_i\mu_i}{\hbar}$ denotes the Rabi frequency  of  each microwave field 
(with $\mathcal{E}_i$ the field amplitude,   $\mu_i$ the transition dipole moment, and $i = 1,2$)
  and   $\Delta_{i}\equiv E_i - E_0 - \hbar\omega_i$  denotes   the single-photon detuning, i.e., the   frequency offset  from each   atomic transition 
 (with $E_{0,i}$ the energies of states $\ket{0,\pm1}$  and $\omega_{i}$  the microwave frequencies). 

The electronic state of the system is described by a density matrix, whose  evolution is governed by the Lindblad master equation~\cite{breuer2002theory,scully1999quantum}. In the Heisenberg picture,  the Lindblad equation of motion for any  observable $\hat{X}$ is
\begin{equation}
\dot{\hat{X}}  \equiv\hat{\mathcal{L}}[\hat{X}]
=  \frac{i}{\hbar}[\hat{H},\hat{X}] +
\sum_j \Gamma_j
\left(
2\hat{L}_j^\dagger \hat{X}\hat{L}_j-\{\hat{L}_j^\dagger\hat{L}_j,\hat{X}\}
\right).
\label{eq:Lindblad}
\end{equation}
The first term represents  Hamiltonian evolution and the remaining terms describe incoherent processes due to the 
interaction with the environment.
We consider   two types of   processes~\cite{breuer2002theory,am2016parameter}: 
($i$) pure dephasing at a rate  $\gamma^{(1,2)}$ (straight dashed arrows) and 
($ii$) downwards or upwards  jumps at   rates  $\kappa_d^{(1,2)}$ or $\kappa_u^{(1,2)}$ respectively  (wiggly dashed arrows).
At thermal equilibrium, the ratio of upwards and downwards transitions is  given by the Boltzmann factor $\kappa_u^{(1,2)}/\kappa_d^{(1,2)} = \exp(\hbar\omega_{1,2}/k_BT)$, where  $k_BT$ is the thermal energy~\cite{am2016parameter}. We assume that the system is   at room temperature ($\approx300K$), where upwards and downwards transition rates are almost equal\footnote{It is    possible to  enhance the  rate of downward transitions using  green light~\cite{doherty2013nitrogen}.
This flexibility allows searching a wider parameter range, as will be explored in future work.}. 
The incoherent  processes are described by~\cite{mathisen2018liouvillian}:
\begin{subequations}
\begin{align}
&\mbox{pure dephasing:}\quad\hat{L}_{1,2} = \ket{\pm1}\bra{\pm1} - \ket{0}\bra{0} \\
&\mbox{up/down  jumps:}\quad\hat{L}_{3,4} = \ket{\pm1}\bra{0}, \quad\hat{L}_{5,6} = \ket{0}\bra{\pm1}
\end{align}
\end{subequations}
with  $\Gamma^{(1,2)} = \gamma^{(1,2)}, \Gamma^{(3,4)} = \kappa_d^{(1,2)}$ and $\Gamma^{(5,6)} =  \kappa_u^{(1,2)}$.

 In order to find EPs in the eigenvalue spectrum of $\hat{\mathcal{L}}$, we rewrite the superoperator equation [\eqref{Lindblad}] in a form that is more convenient for  theoretical investigation~\cite{mathisen2018liouvillian,am2015three}. 
We introduce   a   basis of traceless orthogonal matrices, which spans the space of  density matrices~\cite{byrd2003characterization,goyal2016geometry,kimura2003bloch}.
Specifically, we choose the  eight Gell--Mann matrices ($\hat{\sigma}_1,\hdots,\hat{\sigma}_8$, defined in the appendix), which generalize   Pauli matrices    for three-level systems~\cite{gell2010symmetries}. 
 By applying \eqref{Lindblad} to each Gell--Mann matrix ($\hat{\sigma}_i$) and taking the expectation value of the resulting equation, we obtain 
\begin{align}
\dot{\vec{S}} = \hat{M}(\vec{S} - \vec{S}_\mathrm{eq}).
\label{eq:GM-dynamics}
\end{align}
Here, $\vec{S}$ is an eight-dimensional vector, whose entries are  $S_i = \mathrm{Tr}[\hat{\rho}\cdot\hat{\sigma_i}]$ and $\hat{\rho}$ is the density matrix.
The real parts of the eigenvalues of  $\hat{M}$  are  the relaxation rates of the   eigenmodes, while  $\vec{S}_\mathrm{eq}$ is   the steady state.  Explicit expressions  for $\hat{M}$ and $\vec{S}_\mathrm{eq}$ are given in the appendix [\eqref{Dynamical-matrix} and \eqref{steady-state-vector} respectively]. 
We consider here  room temperatures,  where  $\vec{S}_\mathrm{eq}\approx 0$,
since the rates of incoherent upwards and downwards transitions are equal. 
Left and right eigenvectors  and eigenvalues  of $\hat{M}$ satisfy 
\begin{gather}
\hat{M}\vec{S}_i^R = \lambda_i\vec{S}_i^R \quad,\quad
\hat{M}^T\vec{S}_i^L = \lambda_i\vec{S}_i^L,
\label{eq:normal-modes}
\end{gather}
where the superscript $T$ denotes matrix transposition. The matrix $\hat{M}$ is non-Hermitian  and, hence, can   have EPs. 
At an EP, the left and right eigenvectors of the degenerate eigenmode are  orthogonal~\cite{Moiseyev2011} (i.e., $\vec{S}_i^L\cdot\vec{S}_i^R = 0$).  Therefore,  the condition number, defined as   the secant of the angle between left and right eigenvectors~\cite{trefethen1997numerical},
\begin{gather}
N(\lambda_i) \equiv \frac{1}{|\cos(\theta_i)|}=\frac{|\vec{S}_i^L||\vec{S}_i^R|}{|\vec{S}_i^L\cdot\vec{S}_i^R|},
\label{eq:condition-num}
\end{gather}
 diverges at the EP.

\figrefbegin{level-structure}(b) reveals the location  of EPs in the eigenvalue spectrum of $\hat{\mathcal{L}}$. We scan the parameters of the  right-circularly polarized  field  ($\Omega_1$ and $\Delta_1$) while holding all other parameters fixed (see  caption). At each point in parameter space, we compute the condition numbers of the eigenvalues of  $\hat{M}$  and, then,     plot the maximal  condition number attained. The dark regions in the figure mark the location  of  the  EPs. We  determine the order of the degeneracy by plotting the eigenvectors at selected points along the  dark lines.  We find three lines of EP2s, which intersect at two points of EP3s  (similar to~\citeasnounss{am2015exceptional,am2016parameter}) and an  isolated EP2 at ($\Delta_1^\mathrm{EP},\Omega_1^\mathrm{EP}) \approx (-80,225)$ kHz.  The same   system  can  be used for finding  fourth- and fifth-order EPs, as we show in the appendix.

Next, we demonstrate   swapping  of the instantaneous eigenvalues  along loops that encircle  the isolated EP2. 
\figrefbegin{riemann-sheets} shows  surfaces of the real and imaginary parts of the eigenvalues of $\hat{M}$ [\eqref{Dynamical-matrix}] as a function of  $\Delta_1$ and $\Omega_1$. 
In the shown  parameter regime,  $\hat{M}$ has four simple eigenvalues ($\lambda_{3,4,5,6}$) and two complex-conjugate  pairs of  eigenvalues ($\lambda_{1,2}$ and $\lambda_{7,8}$) that coalesce at  the  isolated EP2. 
We  choose a  loop   that encircles the EP2  and, then,  compute the  eigenvalues at each point along the path. (The path  details  are given in the appendix). 
As expected,  the simple  eigenvalues  return to the starting point   after the loop  (dashed lines), while the remaining eigenstates swap   (arrowed lines). That is,   $\ket{1}\leftrightarrow\ket{2}$ and $\ket{7}\leftrightarrow\ket{8}$; the red-rimmed points  (associated with  $\lambda_{1,8}$) are transferred  into  cyan-rimmed points (associated with $\lambda_{2,7}$) after the cycle  and vice versa.

Having shown that  the   eigenmodes of $\hat{M}$  swap when encircling the   EP2, let us adopt  a more  intuitive picture and transform  the Gell--man vectors into density matrices.  For three-level systems, the transformation  is~\cite{kimura2003bloch},
\begin{align}
\hat{\rho} = \frac{1}{3}\left(\hat{\mathbb{1}} + \sqrt{3}\sum_{i = 1}^8\hat{\sigma_i}S_i\right),
\label{eq:GM2rho}
\end{align}
 where $\hat{\sigma}_i$ are the Gell--Mann matrices (see appendix).  
In order to conserve   the probabilistic interpretation of $\hat{\rho}$, $\vec{S}$ must be real~\cite{byrd2003characterization,goyal2016geometry,kimura2003bloch}.
Consequently,  complex eigenmodes (such as  $\vec{S}_{1,2}^R$)  cannot be used  as inputs for the switch and, instead, we initialize it in symmetric superpositions of      complex-conjugate  vectors, i.e.,   $\vec{S}^{(1)}_\mathrm{in} \propto \vec{S}^R_1 + \vec{S}_8^R$ and $\vec{S}^{(2)}_\mathrm{in} \propto \vec{S}^R_2 + \vec{S}_7^R$. 
Positivity of $\hat{\rho}$  implies that the length of $\vec{S}$ needs to be smaller than a critical value~\cite{kimura2003bloch}. To satisfy this condition, we normalize  $\vec{S}^{(1)}_\mathrm{in}$ and  $\vec{S}^{(2)}_\mathrm{in}$ accordingly. 
Last, we use \eqref{GM2rho} to transform the   Bloch   vectors, $\vec{S}^{(1,2)}_\mathrm{in}$,  into density matrices,  $\hat{\rho}^{(1,2)}_\mathrm{in}$.

Next, we simulate the  evolution of the system.  We initialize the system in either $\hat{\rho}^{(1)}_\mathrm{in}$ or  $\hat{\rho}^{(2)}_\mathrm{in}$ and  solve  \eqref{GM-dynamics}    via the standard  Runge-Kutta  method~\cite{trefethen1997numerical}. 
 \figrefbegin{topo-switch} summarizes our  main result: The initial state $\hat{\rho}^{(1)}_\mathrm{in}$  is adiabatically transported
  into $\hat{\rho}^{(2)}_\mathrm{in}$  when the  parameters are varied in a \emph{clockwise} manner around the EP  [panels (a--b)].   
This happens because $\hat{\rho}^{(1)}_\mathrm{in}$ is the least-decaying eigenstate for a clockwise loop, which guarantees that the probability for quantum jumps into different eigenstates is negligible. 
 Conversely,  $\hat{\rho}^{(2)}_\mathrm{in}$ is adiabatically transported  into $\hat{\rho}^{(1)}_\mathrm{in}$ only after a \emph{counterclockwise} loop around the EP [panels (d--e)].
More details about the dynamics are given in the appendix (\figref{mixed-state-all-options}).  
 Panels (c,f) show  the  projections of the initial and final Gell--Mann vectors  on the  eigenvectors  at the beginning of the loop, reaffirming that $\vec{S}_\mathrm{in}^{(1)}\!\!\rightarrow\!\!\vec{S}_\mathrm{in}^{(2)}$ after a clockwise loop while $\vec{S}_\mathrm{in}^{(2)}\!\!\rightarrow\!\!\vec{S}_\mathrm{in}^{(1)}$ after a counterclockwise loop. 

In order for the system to evolve adiabatically, the sweep rates of $\Delta_1$ and $\Omega_1$ need to be small compared to the ``energy gap''~\cite{bohm2012quantum} (i.e., the distance between the complex eigenvalues). It implies that the cycle  needs to be  long  compared to the dephasing time (e.g., we chose  $T = 15/\gamma^{(1)}$). 
Consequently, the final states, $\vec{S}_\mathrm{out}^{(1,2)}$,  approach zero  [since $\vec{S}_\mathrm{eq} \approx 0$]. From \eqref{GM2rho}, one learns that when $|\vec{S}|\ll1$,  $\hat{\rho}$ is  nearly diagonal, and it is hard to  distinguish the  final states
 from   diagonal  matrices.  For visual clarity,  in \figref{topo-switch},  we set the diagonal elements of $\hat{\rho}$ to zero, and show only  the moduli of the off-diagonal elements. Since these terms are  very small  [$\mathcal{O}(10^{-4})$], it is challenging to distinguish between the final states. 
  Possible approaches for fighting decoherence are discussed in the concluding paragraph. 

Another challenge  arises from the fact that the  input states, $\hat{\rho}^{(1,2)}_\mathrm{in}$,  are mixed. Experimentally, preparing  pure states is a standard procedure~\cite{doherty2013nitrogen} but the preparation of mixed states is  more challenging. One approach for  overcoming  this problem  is to prepare the system in pure states  in the vicinity of $\hat{\rho}^{(1,2)}_\mathrm{in}$. The problem of finding the nearest pure state to a given density matrix  is equivalent to  finding the best rank-one approximation for that matrix, which can be  solved by exploiting the singular value decomposition~\cite{markovsky2008structured,abdi2010principal}. Unfortunately, when applying this algorithm to  $\hat{\rho}^{(1,2)}_\mathrm{in}$, one obtains pure states that have significant population in undesired states\footnote{The  pure states have significant population in  $\vec{S}^R_5$, 
which is the steady state  since $|\mathrm{Re}[\lambda_5]|<|\mathrm{Re}[\lambda_j]|$ for all $j\neq5$ [see \figref{riemann-sheets}(a)]. Since the cycle  is much longer than the coherence lifetime, the final state is almost precisely $\vec{S}^R_5$, and the swapping of  the  partial populations of $\vec{S}_\mathrm{in}^{(1)}$ and $\vec{S}_\mathrm{in}^{(2)}$ becomes  unmeasurable.},  which deteriorate the performance of the switch. 
Alternatively, one could use quantum control to find protocols for preparing arbitrary mixed states~\cite{bartana1997laser,jirari2005optimal,roloff2009optimal,yan1993optical}.
However,  experimental implementation  of such    protocols may be very difficult. These direction will be explored in future work.

To summarize, we presented a protocol for achieving  robust  mode conversion  with  NV centers by using EPs. 
Our work  generalizes  the existing theory  of EP-based  switches in two aspects: 
1. By including  mixed states, which are necessary for describing  most existing  platforms for quantum information processing, and 2. By treating  multilevel systems, generalizing previous  work that focused  on two-level systems.  We find that EP-based  mode switches in  Lindbladian systems require  at least three electronic levels, and  that one could force high-order EPs   by carefully tuning several control  parameters [see ~\figref{xenon}].  
Our analysis raises two  challenges for experimental demonstration of this  protocol: mixed-state preparation and decoherence.  The former challenge is technical, and several ways for addressing it  are mentioned above. The latter---fighting decoherence---is more fundamental. Quite generally, there is an  incompatibility between  adiabaticity  (which implies  slow evolution) and maintaining coherence  (which requires, in turn,  fast  evolution)~\cite{miniatura1990geometrical}.   Adiabaticity requires that the parameter sweep rate  should be smaller than the energy gap~\cite{born1928beweis} which, in our case, is on the order of the decoherence rate. Unfortunately, it  implies that whenever the evolution period is long enough to enable adiabatic evolution, the signal   degrades substantially. This shortcoming can be remedied  by using modified protocols.  For example, one could introduce additional lasers that  actively restore  the lost coherence during the evolution. Such ``cycling transitions'' exist in NV centers, and require cold-temperature conditions~\cite{togan2011laser}. Another possibility is to use  the notion of PT symmetry, which implies that under some  appropriate conditions, a PT-symmetric non-Hermitian system   evolves without dissipation~\cite{bender1999pt,ruter2010observation}. PT-symmetric evolution was recently observed in NV centers~\cite{wu2019observation},   and this approach can be extended  to  design a PT-symmetric mode switch.  Such  protocols  are expected to significantly improve the performance of the switch, and will be addressed in future work.

\section*{Acknowledgment}
The authors would like to thank  Ronnie Kosloff, Saar Rahav,    Alexei Mailybaev,  Yaniv Kurman, Sigal Wolf,  and Idan Meirzada for helpful discussions.  AP is supported   by an Aly Kaufman Fellowship at the Technion and  by The Center for Absorption in Science, Ministry of Immigrant Absorption, State of Israel.  NM acknowledges the financial support of I-Core: The Israeli Excellence Center ``Circle of Light,'' and of the Israel Science Foundation (Grant No. 1530/15). NB acknowledges support from the European Union (ERC StG, MetaboliQs), the CIFAR Azrieli Global Scholars program, the Ministry of Science and Technology, Israel, and the Israel Science Foundation (Grant No. 750/14).

\section{Appendix}

 
  \begin{figure*}[t]
  \begin{center}
    \includegraphics[scale=0.3]{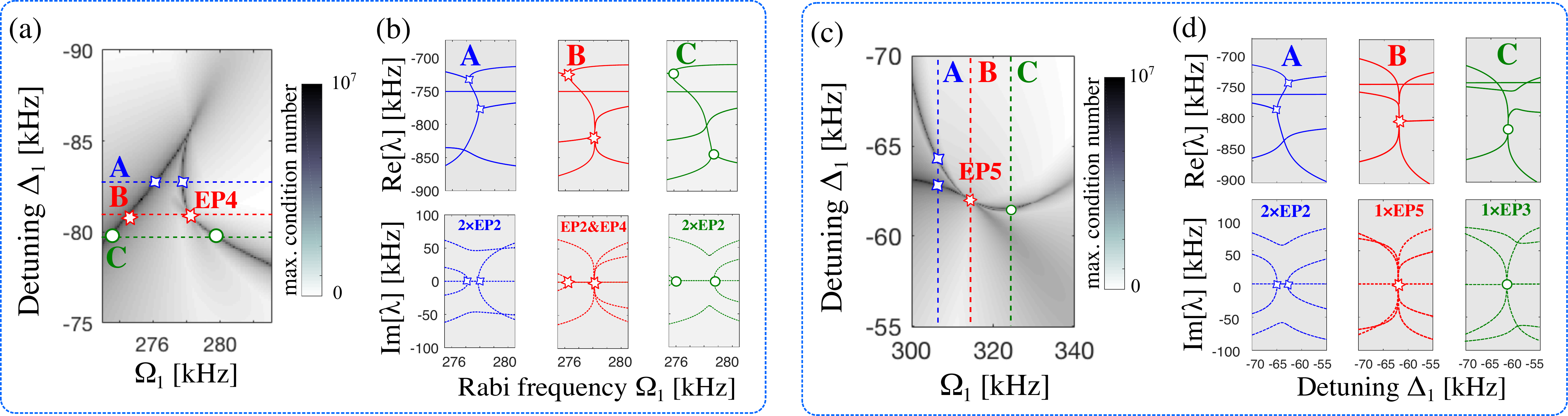}
  \end{center}
  \caption{\textbf{Higher-order EPs,} found by searching the four-dimensional parameter space
  spanned by the two Rabi frequencies $\Omega_{1,2}$ and frequency offsets $\Delta_{1,2}$ [see \figref{level-structure}(a)]. 
   (a) A fourth-order degeneracy is found at $\Delta_1 = -80.8, \Omega_1 = 278.2, \Delta_2 = 44.3, \Omega_2 =  -445$ (in units of kHz). 
   (b) The plots show real  and imaginary parts of the eigenvalues (top and bottom plots) as a function of $\Omega_1$, at three values of $\Delta_1$, above(A) at (B) and below the EP (C). The plots demonstrate that two pairs of second-order EPs (EP2s) merge into an EP4. 
  (c) A fifth-order degeneracy is found at $\Delta_1 = -62, \Omega_1 = 314, \Delta_2 = 58, \Omega_2 = 436$. 
  (d)Similar to (b), the plots show the real and imaginary parts of the eigenvalues along the three cuts (labeled  A--C) in panel (c). The plots demonstrate that two pairs of EP2s merge with an EP3 to form the fifth-order EP. 
  }
  \label{fig:xenon}
\end{figure*}
 
\subsection{Adiabatic theorem for non-Hermitian systems}
\renewcommand{\theequation}{A\arabic{equation}}
\setcounter{equation}{0}
In this appendix, we sketch the proof of the adiabatic theorem for non-Hermitian systems following~\citeasnoun{nenciu1992adiabatic}.
Let us investigate  the conditions for adiabatic evolution of the normal modes  of a non-Hermitian time-dependent operator, $\hat{M}(\varepsilon t)$ (with $\varepsilon>0$). We introduce  a new  variable, $s = \varepsilon t$, and  consider the limit of $\varepsilon\rightarrow0$ while $s$ is held  fixed and finite. By invoking the normal-mode expansion of $\hat{M}(s)$, one can write 
\begin{gather}
\hat{M}(s) = \sum_{j = 1}^n \lambda_j(s) \ket{\psi^R_j(s)}\bra{\psi^L_j(s)},
\end{gather}
where  the time-dependent right- and lef-eigenvectors and corresponding eigenvalues are defined in~\eqref{normal-modes} in the main text.  The normal modes satisfy the biorthogonality condition 
$\bracket{\psi^L_i(s)}{\psi^R_j(s)}=\delta_{ij}$.
Let $\ket{\psi(s)}$ be a solution of the  differential equation 
\begin{gather}
\varepsilon \tfrac{\partial }{\partial s}  \ket{\psi(s)}= \hat{M}(s)\ket{\psi(s)}.
\label{eq:ODE}
\end{gather} 
Substituting the following ansatz 
\begin{gather}
\ket{\psi(s)} = \sum_j a_j(s) \exp\left(-\frac{1}{\varepsilon}\int_{s_0}^s \lambda_j(u)du\right)
\end{gather}
into \eqref{ODE} and using the biorthogonality condition,  one obtains
\begin{align}
&\partial_s{a}_\ell(s) + \bracket{\psi^L_\ell(s)}{\partial_s \psi^R_\ell(s)}{a}_\ell(s)=\nonumber\\
&-\sum_{j\neq\ell}
\bracket{\psi^L_\ell(s)}{\partial_s\psi^R_j(s)}a_j(s)
\Scale[0.9]{
\exp\left(
-\frac{1}{\varepsilon}
\int_{s_0}^s
\left[\lambda_j(u)-\lambda_\ell(u)
\right]du
\right)}
\label{eq:adiabatic-conditions}
\end{align}
From \eqref{adiabatic-conditions}, one can easily read the conditions for adiabatic evolution. 
For Hermitian systems [where $\hat{M}(s) = \hat{M}^\dagger(s)$], the eigenvalues [$\lambda_j(s)$] are real and the right-hand side  of \eqref{adiabatic-conditions} contains only rapidly oscillating terms, which  average to zero in the limit of $\varepsilon\rightarrow0$.
Conversely,  when $\hat{M}(s)$ is non-Hermitian, the right-hand side of \eqref{adiabatic-conditions} can be neglected only when 
\begin{align}
\Gamma_{\ell j}(s) = 
 \int_{s_0}^s
\mathrm{Re}\left[\lambda_j(u)-\lambda_\ell(u)
\right]du >0
\end{align}
 for all $j\neq\ell$. 
To summarize,  the  condition for adiabatic evolution  comes from requiring   that the probability  to  jump from state  $\ell$ to  $j$   is small. In contrast to Hermitian systems,  where the probability for quantum jumps   oscillates in time, in  non-Hermitian systems, it also  contains exponentially growing or decaying factors, and can be small only if  $\Gamma_{\ell j}(t)>0$ for all $j\neq\ell$.

\subsection{Vectorizing the Lindblad master equation}
\renewcommand{\theequation}{B\arabic{equation}}
\setcounter{equation}{0}

In this appendix we derive \eqref{GM-dynamics} from the main text and present explicit expressions for the dynamical matrix $\hat{M}$ and the steady-state Gell--Mann vector $\vec{S}_\mathrm{eq}$.  A similar formulation was introduced in~\cite{mathisen2018liouvillian} to analyze the stimulated Raman adiabatic passage (STIRAP) method~\cite{vitanov2001laser}.
To this end, we introduce the set of  Gell--Mann matrices~\cite{gell2010symmetries}:
\begin{gather}
\hat{\sigma}_1 =  \left( \begin{array}{ccc}
0    & 1 & 0 \\
1 & 0 & 0\\
0&0&0\end{array} \right)
\quad
\hat{\sigma}_2 =  \left( \begin{array}{ccc}
0    & -i & 0 \\
i & 0 & 0\\
0&0&0\end{array} \right)
\nonumber\\
\hat{\sigma}_3=  \left( \begin{array}{ccc}
1    & 0 & 0 \\
0 & -1 & 0\\
0&0&0\end{array} \right)
\quad
\hat{\sigma}_4 =  \left( \begin{array}{ccc}
0  & 0 & 1 \\
0 & 0 & 0\\
1 & 0 & 0 \end{array} \right)\nonumber\\
\hat{\sigma}_5 = 
 \left( \begin{array}{ccc}
0  & 0 & -i \\
0 & 0 & 0\\
i & 0 & 0 \end{array} \right)
\quad 
\hat{\sigma}_6 =  \left( \begin{array}{ccc}
0    & 0 & 0 \\
0 & 0 & 1\\
0&1&0\end{array} \right)
\nonumber\\
\hat{\sigma}_7 =  \left( \begin{array}{ccc}
0    & 0 & 0 \\
0 & 0 & -i\\
0&i&0\end{array} \right)
\quad
\hat{\sigma}_8 = 
\frac{1}{\sqrt{3}}
\left( \begin{array}{ccc}
1    & 0 & 0 \\
0 & 1 & 0\\
0&0&-2\end{array} \right)
\end{gather}
We apply  the Lindblad master equation [\eqref{Lindblad}] to each Gell-Mann operator ($\hat{\sigma_i}$) and compute  the expectation value, obtaining \eqref{GM-dynamics} from the main text, with 
\begin{align}
&\Scale[0.8]{\hat{M} \equiv }\nonumber\\
&\Scale[0.8]{
 \left( \begin{array}{cccccccc}
-\gamma_{11}  & \Delta_1 & 0 & 0 & 0 & 0 & \Omega_2 & 0\\
-\Delta_1  & -\gamma_{22} & -2\Omega_1 & 0 & 0 & \Omega_2 & 0 & 0\\
0  & 2\Omega_1 & -\gamma_{33} & 0 & \Omega_2 & 0 & 0 &  \gamma_{38}\\
0  & 0 & 0 & -\gamma_{44} & \Delta_2 & 0 & -\Omega_1 & 0\\
0  & 0 & -\Omega_2 & -\Delta_2 &  -\gamma_{55} & \Omega_1 & 0 & -\sqrt{3}\Omega_2 \\
0  & -\Omega_2 & 0 & 0 & -\Omega_1 & -\gamma_{66} & -(\Delta_1 - \Delta_2) & 0\\
-\Omega_2  & 0 & 0 & \Omega_1 & 0 & (\Delta_1 - \Delta_2) & -\gamma_{77} & 0\\
0  & 0 & \gamma_{83} & 0 & \sqrt{3}\Omega_2 & 0 & 0 & -\gamma_{88}\end{array} \right)},
\label{eq:Dynamical-matrix}
\end{align}
where we introduced the notation
\begin{subequations}
\begin{align}
&\gamma_{11} = \gamma_{22} \equiv \left(\tfrac{\gamma_2}{8}  + \tfrac{\gamma_1 + \kappa_d^{(1)}+\kappa_u^{(1)}+\kappa_u^{(2)}}{2}\right)\\
&\gamma_{33} \equiv \tfrac{\kappa_u^{(2)}}{2} + \kappa_d^{(1)} + \kappa_u^{(1)}\\
&\gamma_{38} \equiv \left(
\kappa_d^{(1)} - \kappa_d^{(2)}-\kappa_u^{(1)}-\tfrac{\kappa_u^{(2)}}{3}
\right)\tfrac{1}{\sqrt{3}}\\
&\gamma_{44}  = \gamma_{55} \equiv \tfrac{\gamma_1}{8} + 
\tfrac{\gamma_2 + \kappa_u^{(1)}+ \kappa_d^{(2)} + \kappa_u^{(2)}}{2}\\
&\gamma_{66} = \gamma_{77} \equiv
\tfrac{\gamma_1 + \gamma_2}{8} + \tfrac{\kappa_d^{(1)}+\kappa_d^{(2)}}{2}\\
&\gamma_{83} \equiv  -\tfrac{\sqrt{3}}{2}\kappa_u^{(2)}\\
&\gamma_{88} \equiv \tfrac{\kappa_u^{(2)}}{2}+\kappa_d^{(2)}.
\end{align}
\end{subequations}
The steady-state vector is given by 
\begin{align}
-(\hat{M}^{-1}\vec{S}_\mathrm{eq})^T = 
 \left( \begin{array}{cccccccc}
0 &0 & \tfrac{\Delta\kappa_2}{3}+\tfrac{2\Delta\kappa_1}{3} & 0 & 0 & 0 & 0& \tfrac{\Delta\kappa_2}{\sqrt{3}}\end{array} \right),
\label{eq:steady-state-vector}
\end{align}
where we define the difference between upwards and downwards jumps as
\begin{align}
\Delta\kappa_i\equiv \kappa_d^{(i)} - \kappa_u^{(i)}.
\label{eq:delta-kappa}
\end{align}
When the upwards and downwards rates are equal (i.e., in the high-temperature limit), the steady-state vector vanishes  and the Gell--Mann vector will approach the origin  at asymptotically large evolution times.

  \begin{figure*}[t]
  \begin{center}
    \includegraphics[scale=0.5 ]{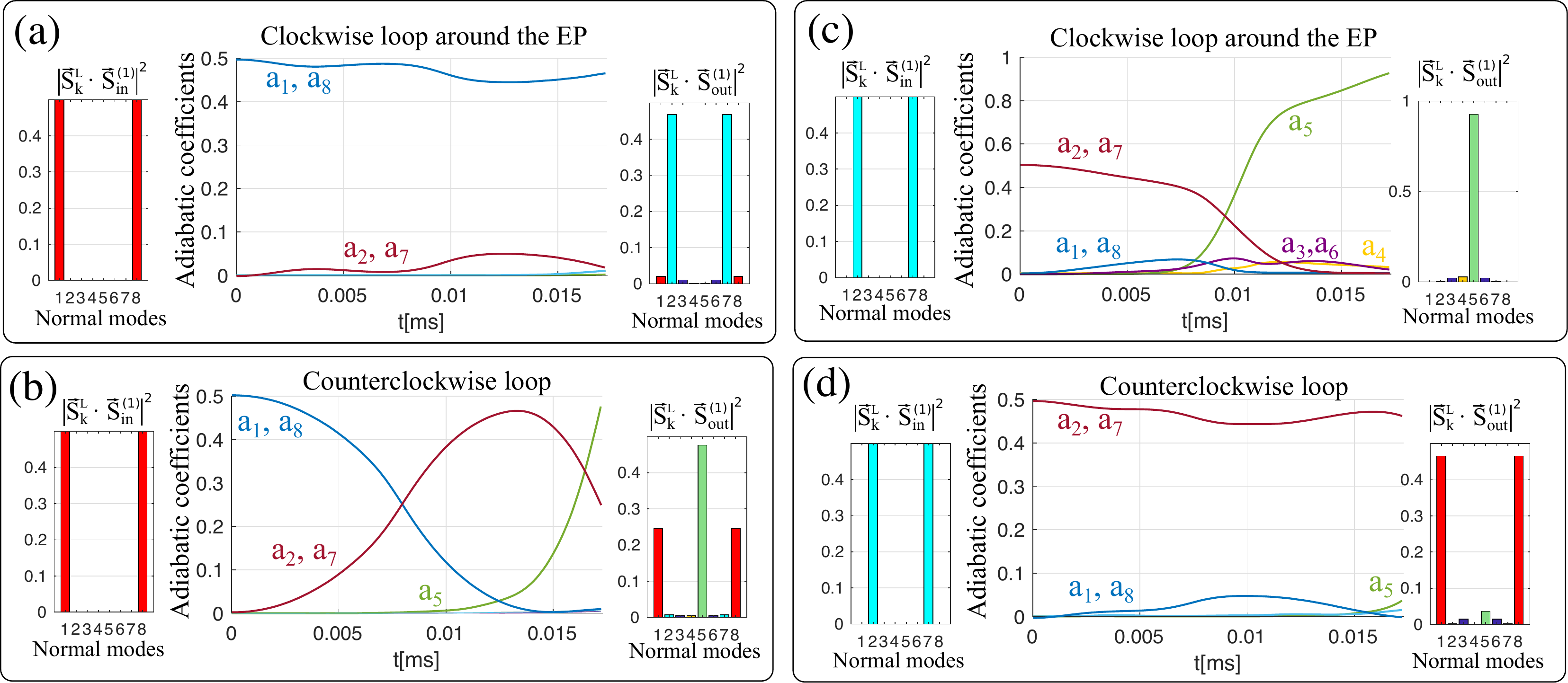}
  \end{center}
  \caption{\textbf{Dynamically encircling an isolated EP}. The middle panels (a--d) show the evolution  of the adiabatic coefficients, $|a_i(t)|^2 = |\vec{S}_i^L(t)\cdot\vec{S}(t)|^2$ [\eqref{adiabatic-coefs}] during the loop. 
The side panels in (a--d) show the normalized projections of the initial and final states onto the instantaneous states at the beginning of the loop, $|\vec{S}_i^L(0)\cdot\vec{S}_{\mathrm{in/out}}^{(1,2)}|^2$.
(a)  When initialized in state $\vec{S}^{(1)}_\mathrm{in}$ and evolved in a clockwise manner, the coefficients $a_1$ and  $a_8$ are predominant throughout the evolution, which implies that the state evolves adiabatically, hence the final state is the coalescing partner state, $\vec{S}^{(2)}_\mathrm{in}$.  (b) When reversing the direction of the loop, the system does not evolve adiabatically and the final state is $\vec{S}_5^R$. Plots (c--d) show the situation when the system is initialized in $\vec{S}^{(2)}_\mathrm{in}$, where only a counterclockwise loop enables adiabatic mode swapping. The parameters are the same as in \figref{topo-switch} from the main text.
   }
  \label{fig:mixed-state-all-options}
\end{figure*}

\subsection{High-order EPs in a three-level system}
\renewcommand{\theequation}{C\arabic{equation}}
\setcounter{equation}{0}
 
In the main text, we show the existence of second- and third-order EPs in the ground-state manifold of the NV center.
 Here, we show that  it is also possible to find high-order EPs in this system. For example, by searching the four-dimensional space spanned by  $\Omega_{1,2}$ and  $\Delta_{1,2}$ (see text), one can induce fourth- and  fifth-order EPs, as shown in  \figref{xenon}. More generally,  the density matrix of an $N$-level system has $N^2 - 1$ real degrees of freedom. So  one could potentially find eighth-order EPs  in this system, but that would require searching a higher-dimensional  parameter space. In order to find a real degenerate eigenvalue of degree $M$, one needs $M-1$ real parameters [to satisfy $\lambda_1(\vec{p}) = \hdots = \lambda_M(\vec{p})$]. In order to find  a complex degenerate eigenvalue   of degree  $M$, one needs $2(M-1)$ real parameters.  We find high-order EPs by using the  algorithm from~\citeasnoun{mailybaev2006computation}, which   exploits  versal deformation theory for finding  EPs of a given order. 

\subsection{Dynamically encircling an isolated EP}
\renewcommand{\theequation}{E\arabic{equation}}
\setcounter{equation}{0}

In this appendix, we provide additional information about the evolution of the system during the loop in parameter space  when encircling the EP in both directions (i.e., clockwise and counterclockwise).  We choose an elliptic path of the form
\begin{gather}
 \Delta_1(t) = \Delta_\mathrm{EP} + R_\Delta \cos(2\pi t/T+\phi), \nonumber\\
   \Omega_1(t) = \Omega_\mathrm{EP} + R_\Omega \sin(2\pi t/T+\phi).
\end{gather}
In \figref{riemann-sheets},   we use $\Delta_\mathrm{EP} = -80, \Omega_\mathrm{EP} = 295,  R_\Delta = 100$, and 
$R_\Omega = 30$  (all in kHz), while in   \figref{topo-switch}, the radii  are  $R_\Delta = 260$  and $R_\Omega = 125$. The phase is $\phi = 0.39\pi$ and the period is  $T = 15/\gamma^{(1)}$.

When initialized in $\vec{S}^{(1)}_\mathrm{in}$, the system evolves adiabatically only when the EP is encircled in a clockwise manner;  the opposite is true for the second initial state, $\vec{S}^{(2)}_\mathrm{in}$.
At each moment along  the evolution,  the instantaneous normal modes form a complete basis of the Hilbert space. That is, the  state vector at time $t$  can be written in the form
\begin{equation}
\vec{S}(t) = \sum_{ i = 1}^8a_i(t)\vec{S}_i^R(t)
\label{eq:adiabatic-coefs}
\end{equation}
where  $a_i(t)$ are called  ``the adiabatic coefficients''. We compute them be projecting the state vector, $\vec{S}(t)$,  onto the instantaneous basis states, $\vec{S}_i^L(t)$. The system evolves adiabatically if it stays in the same instantaneous eigenstates throughout its evolution.   It is important to emphasize that the instantaneous eigenstates themselves  swap at the end of the loop; that is, if the system starts and ends with predominant coefficients $a_1$ and $a_8$, it means that the state swapped because $\vec{S}_1^R(t=T) = \vec{S}_2^R(t=0)$ and $\vec{S}_8^R(t=T) = \vec{S}_7^R(t=0)$. \figrefbegin{mixed-state-all-options} shows the evolution of the adiabatic coefficients during the loop [middle panels in (a--d)] and the projections of the input and output states on the instantaneous eigenstates (i.e., the normal modes) at the beginning of the evolution. More details are given in the caption.

%


\end{document}